%% file: main.tex
\documentclass[runningheads]{llncs}
\usepackage[T1]{fontenc}
\usepackage{graphicx}
\usepackage{tablefootnote}
\begin{document}
\title{No Data? No Problem: Synthesizing Security Graphs for Better Intrusion Detection}
\titlerunning{Synthesizing Security Graphs for Better Intrusion Detection} 

\author{Yi Huang\inst{1} \and
Shaofei Li\inst{1} \and
Yao Guo\inst{1} \and
Xiangqun Chen\inst{1} \and
Wajih Ul Hassan\inst{2} \and
Ding Li\inst{1}
}

\authorrunning{Y. Huang et al.}

\institute{
Key Laboratory of High-Confidence Software Technologies (MOE), School of Computer Science, Peking University, Beijing, China\\
\email{yihuang@stu.pku.edu.cn, \{lishaofei, yaoguo, cherry, ding\_li\}@pku.edu.cn}
\and
Department of Computer Science, University of Virginia, Charlottesville, VA, USA\\
\email{hassan@virginia.edu}
}

\maketitle

\begin{abstract}
  Provenance graph analysis plays a vital role in intrusion detection, particularly against Advanced Persistent Threats (APTs), by exposing complex attack patterns. While recent systems combine graph neural networks (GNNs) with natural language processing (NLP) to capture structural and semantic features, their effectiveness is limited by class imbalance in real-world data. To address this, we introduce \Sys, a novel hybrid provenance graph synthesis framework, which comprises three  components: (1) graph structure synthesis via heterogeneous graph generation models, (2) textual attribute synthesis via fine-tuned Large Language Models (LLMs), and (3) five-dimensional fidelity evaluation. Experiments on six benchmark datasets demonstrate that \Sys consistently produces higher-fidelity graphs across the five evaluation dimensions compared to four strong baselines. To further demonstrate the practical utility of \Sys, we utilize the synthesized graphs to augment training datasets for downstream APT detection models. The results show that \Sys effectively mitigates data imbalance, improving normalized entropy by up to 0.35 in absolute terms, and enhances the generalizability of downstream detection models, yielding an absolute increase of up to 0.38 in balanced accuracy.
\end{abstract}

\input{body}

\input{discussion}

\section{Conclusion}
This paper proposes \Sys, a novel provenance graph generation framework consisting of three components: graph structure synthesis via heterogeneous graph generation models, textual attribute synthesis via fine-tuned LLMs, and five-dimensional fidelity evaluation. Compared to provenance data synthesized directly by four state-of-the-art LLMs, \Sys consistently produces graphs with higher fidelity. When integrated into APT detection training sets, \Sys-generated graphs effectively mitigate class imbalance and yield improved accuracy.

\begin{credits}
\subsubsection{\ackname}
This work was supported by the National Natural Science Foundation of China
under Grant No.~U25A6024, and by the Fundamental and Interdisciplinary Disciplines
Breakthrough Plan of the Ministry of Education of China under Grant No.~JYB2025XDXM108.


\end{credits}

\bibliographystyle{splncs04}
\bibliography{ref}

\appendix
\input{appendix}

\end{document}

%% file: body.tex
\section{Introduction}
Data provenance~\cite{inam2023sok} records system activities for cybersecurity attack tracing~\cite{wang2020you}. Recent deep learning-based \ac{pids}~\cite{han2020unicorn,wang2022threatrace,zengy2022shadewatcher,rehman2024flash,li2023nodlink} effectively capture system complexity and detect sophisticated attacks like zero-day and \ac{apt} attacks. These systems typically rely on unsupervised or self-supervised learning to model normal system behaviors. However, their effectiveness is constrained by training dataset quality and diversity~\cite{11023260}. These systems assume training data adequately represents diverse runtime configurations, workload patterns, and benign activities. However, this assumption is often violated in real-world scenarios where publicly available datasets exhibit significant limitations.

Datasets used for training and evaluating learning-based \ac{pids} often suffer from \textbf{data imbalance}, where certain system behaviors are overrepresented while others are severely underrepresented. These datasets simulate cyber attacks primarily from controlled experimental environments such as the DARPA TC and OpTC datasets~\cite{darpa-E3}. Prior studies show that benign activities in these datasets predominantly originate from automated workload generators, producing highly monotonous and predictable background behaviors~\cite{11023260}. Consequently, models trained on such data tend to overfit to repetitive patterns and fail to generalize to diverse, unpredictable activities in production environments. 
This data imbalance hampers the generalizability of machine learning-based \ac{pids}, limiting their effectiveness in diverse real-world deployments~\cite{zhou2023improving,chawla2002smote,zhao2021graphsmote,anjum2021analyzing,11023260}.

To synthesize provenance graphs for data augmentation, our key insight is that provenance graphs encode two complementary types of information: \textit{structural information} capturing relationships and interaction patterns among system entities, and \textit{semantic information} describing entity attributes and identities. Both are critical for learning-based intrusion detection, as malicious activities are jointly determined by structural patterns (e.g., abnormal execution chains) and semantic attributes (e.g., suspicious file paths or command-line arguments)~\cite{li2023nodlink,rehman2024flash}. However, existing synthesis approaches fail to handle this dual nature. Graph synthesis methods~\cite{you2018graphrnn,goyal2020graphgen,de2018molgan} focus on generating topologies and categorical labels while neglecting rich textual attributes essential for intrusion detection. Conversely, LLM-based text generation excels at producing semantically rich log entries but fails to preserve graph structure when converted to provenance graphs, resulting in structurally invalid or semantically inconsistent connections. Therefore, effective synthesis must jointly model both structural connectivity and semantic attributes to generate high-fidelity provenance graphs for robust detection systems.

To address these limitations, we propose \Sys, a hybrid provenance graph synthesis framework integrating graph generation models and large language models. \Sys jointly models both structural and semantic information through a two-phase pipeline. In the \textit{graph structural synthesis} phase, a heterogeneous graph generation model constructs topologies with accurate node and edge type distributions, capturing structural patterns from real provenance data. In the \textit{textual attribute synthesis} phase, a fine-tuned LLM generates realistic textual attributes for each node, ensuring semantic consistency with the graph structure. However, realizing this framework introduces three principal technical challenges:

\noindent \textbf{Scalability to provenance graphs.} Provenance graphs often contain millions of nodes and edges, whereas most existing heterogeneous graph generation models are trained on graphs with fewer than 100 nodes. To address this scalability challenge, we adopt time-bounded bidirectional subgraph sampling to extract representative subgraphs. Unlike generic sampling methods such as node sampling or random walks, this strategy preserves both temporal locality and directed dependency structure by expanding from a seed node along both incoming and outgoing edges within a bounded time window. We then train an enhanced GraphGen~\cite{goyal2020graphgen} on the sampled subgraphs to jointly model node types, edge types, and structural connectivity. In particular, we incorporate a direction-aware DFS encoding to represent information flow and causal dependencies, and a schema-constrained decoder to enforce valid relations.

\noindent \textbf{Textual attributes synthesis.} Although LLMs demonstrate strong text generation capabilities, they struggle with graph structure comprehension~\cite{jin2024large} and lack domain-specific provenance knowledge. Consequently, directly prompting LLMs often produces generic or unrealistic outputs. To address this, we first design a \ac{sft} strategy adapting LLMs to the provenance domain. Then we convert provenance graphs into \ac{dfs} sequences, transforming graph structure into a linear format suitable for LLM processing while preserving connectivity and relationships.

\noindent \textbf{Evaluation framework gap.} A critical challenge in provenance graph synthesis is the absence of comprehensive evaluation frameworks. Existing methods~\cite{you2018graphrnn,li2018learning,goyal2020graphgen}, focusing on topological properties are inadequate for provenance graphs requiring verification of both structural validity and semantic plausibility. To address this gap, we propose a novel five-dimensional fidelity evaluation framework, including \textit{structural}, \textit{textual}, \textit{temporal}, \textit{embedding}, and \textit{semantic} metrics. This framework enables rigorous assessment of synthesized provenance graphs and addresses a critical gap in the graph generation literature.

In the experiments, we evaluate the fidelity of provenance data synthesized by \Sys following our proposed evaluation framework. We compare \Sys against four advanced LLMs (GPT-5.1~\cite{openai-gpt51}, Claude-sonnet-4.5~\cite{claude}, DeepSeek-v3.2~\cite{liu2025deepseek}, and Qwen3-max~\cite{yang2025qwen3}) on six widely used provenance datasets. The results show that \Sys achieves higher fidelity than directly using these LLMs from all five evaluation perspectives.
Moreover, data imbalance is consistently mitigated after incorporating \Sys, improving normalized entropy by up to 35\%. 
We further demonstrate the practical utility of \Sys by augmenting existing datasets with the synthesized graphs for improving four latest downstream APT detection models (NodLink~\cite{li2023nodlink}, Magic~\cite{jia2024magic}, Flash~\cite{rehman2024flash}, Velox~\cite{bilot2025sometimes}). Experimental results demonstrate that detection models trained on \Sys-augmented datasets yield up to a 38\% gain in balanced accuracy, while also improving discriminative reliability and diagnostic utility.

The main contributions of our work are as follows:
\begin{itemize}[topsep=.2ex,itemsep=.2ex,leftmargin=*]
    \item We propose \Sys, a novel hybrid provenance graph synthesis framework that jointly models structural topology via graph generation models and textual semantics via fine-tuned LLMs, enabling the construction of high-fidelity provenance graphs.
    \item We introduce a novel five-dimensional fidelity evaluation framework specifically designed for provenance graph synthesis, encompassing structural, textual, temporal, embedding, and semantic metrics, addressing a critical gap in existing graph generation research.
    \item Through systematic experiments on six benchmark datasets, we demonstrate that \Sys effectively mitigates data imbalance and enhances the generalizability of downstream \ac{apt} detection models.
    \item We open source \Sys at \url{https://github.com/hyiii0204/OpenProvSyn} to facilitate future research in provenance graph synthesis and \ac{pids}.
\end{itemize}

\section{Background and Motivation}

\subsection{Provenance Graph}
\label{sec:background:provenancegraph}
Provenance graphs are constructed from system audit events that record interactions among system entities, such as processes, files, and network connections. \textit{Formally, a provenance graph is a directed heterogeneous graph with textual attributes $G = (V, E)$, where
each node $v_i \in V$ is associated with a type $\tau_i$ and a textual attribute: a name $n_i$, and each edge $e_i \in E$ is associated with a type $l_i$.} Each system audit event is represented as a labeled 5-tuple:
$
e_i = (\tau_i, n_i, \tau_j, n_j, l_i)
$, 
where $\tau_i$ and $\tau_j$ denote the types of the source and target nodes, $n_i$ and $n_j$ denote their names, and $l_i$ denotes the edge type. Edges are directed from the source node to the target node, reflecting the causal relationships between entities.

For node types, the most common node types include \emph{process}, \emph{file}, and \emph{network}.
For node names, process nodes are typically command lines, file nodes are their absolute file paths, and network nodes are their IP addresses. 
For edge types, common process--process interactions include \texttt{CLONE}, \texttt{SENDMSG}, and \texttt{RECVMSG}, which capture process creation and message-based communication.
Common process--file interactions include \texttt{READ}, \texttt{WRITE}, and \texttt{OPEN}, which represent file access and modification behaviors.
Common process--network interactions include \texttt{CONNECT}, \texttt{SENDTO}, and \texttt{RECVFROM}, which describe network connection establishment and data transmission.

\subsection{Class Imbalance}
Class imbalance poses a significant challenge in training effective machine learning models across diverse tasks, including financial fraud detection~\cite{dal2015calibrating,roy2024frauddiffuse}, hate speech identification~\cite{davidson2017automated,caselli2021hatebert}, and rare disease diagnosis~\cite{esteva2017dermatologist,tiu2022expert}. Class imbalance refers to a dataset characteristic where the number of samples across classes is uneven, often with one or more classes dominating the dataset while the remaining classes account for only a small fraction~\cite{japkowicz2002class}.
Previous studies have revealed that class imbalance has a detrimental effect on a model’s classification performance as well as its generalization abilities in various machine learning tasks~\cite{buda2018systematic,johnson2019survey}. When deployed in open-world out-of-distribution (OOD) environments, detection models trained on imbalanced datasets may significantly exhibit performance degradation~\cite{yang2024generalized,he2009learning}. Hence, bridging the class imbalance gap in existing training set is vital for machine learning models to achieve robust results.

\begin{figure}[!t]
    \centering
    \includegraphics[width=0.6\columnwidth]{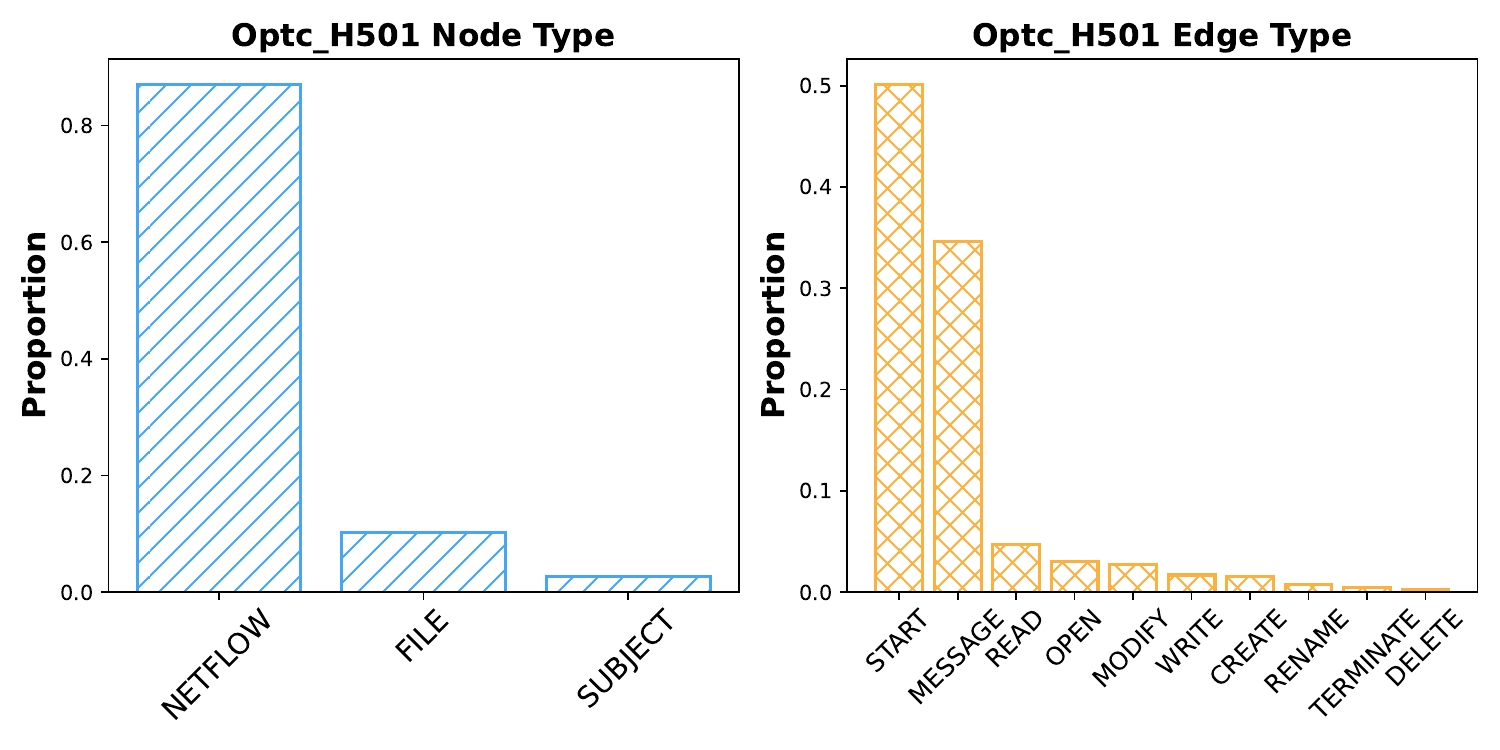}
    \caption{Entity Type and Event Type Distribution in Provenance Dataset OpTC-H501.}
    \label{fig:OpTC_label_stats}
\end{figure}
To further illustrate the class imbalance issue in provenance datasets, we statistically analyze the distributions of entity types and event types across six commonly used public provenance datasets, including Cadets-E3, Theia-E3, ClearScope-E5, Theia-E5, OpTC-H201, and OpTC-H501. Figure \ref{fig:OpTC_label_stats} presents the distribution of OpTC-H501, while the distributions of the other datasets exhibit comparable distributional patterns and are shown in Appendix \ref{sec:entity-event-distribution} (Figure \ref{fig:label_stats}). These distributions exhibit a typical long-tail pattern and reveal substantial data imbalance. For both entity types and event types, a single category accounts for more than 50\% of the instances, exceeding 80\% in some cases. In contrast, most other categories contribute less than 5\%. This imbalance further reflects the uneven distribution of system behaviors in most provenance datasets.

\subsection{Limitations of Existing Methods}

To mitigate class imbalance, conventional techniques such as undersampling~\cite{liu2008exploratory}, oversampling~\cite{mohammed2020machine} and synthetic minority oversampling technique (SMOTE)~\cite{chawla2002smote} have been widely adopted. However, these methods are unsuitable for provenance graphs. Undersampling discards majority-class samples, disrupting causality chains between system entities. Oversampling replicates minority-class instances without introducing diversity, potentially causing overfitting. SMOTE generates synthetic samples through linear interpolation between neighbors in feature space; however, due to overlapping class distributions, it cannot be effectively applied to provenance graphs where structural validity must be preserved.

An alternative approach leverages graph generation models to synthesize provenance graphs by learning underlying topology distributions. While frameworks such as MolGAN~\cite{de2018molgan}, GraphRNN~\cite{you2018graph}, and GraphGen~\cite{goyal2020graphgen} effectively capture structural connectivity, they primarily optimize adjacency patterns and categorical labels while neglecting textual attributes. In contrast, provenance graphs contain rich textual attributes associated with nodes—such as command lines, file paths, and IP addresses—that are crucial for downstream intrusion detection systems~\cite{li2023nodlink,rehman2024flash}. To date, no graph generation model inherently supports joint generation of both graph structure and textual attributes.

Recent advances in LLMs demonstrate strong capabilities in natural language understanding and generation~\cite{achiam2023gpt,claude}, motivating exploration of LLM-based provenance data synthesis. Specifically, LLMs can generate provenance log entries containing rich textual and semantic information. However, when converting LLM-generated logs into graphs and evaluating them using graph-based metrics, their structural fidelity falls significantly short of graphs produced by heterogeneous graph generation models (see Section~\ref{subsubsec:structure-eval}). This indicates that although current LLMs perform well on text synthesis, they lack strong capability for synthesizing accurate graph topology and structure.

\begin{figure*}[t!] 
    \centering
    \includegraphics[width=\textwidth]{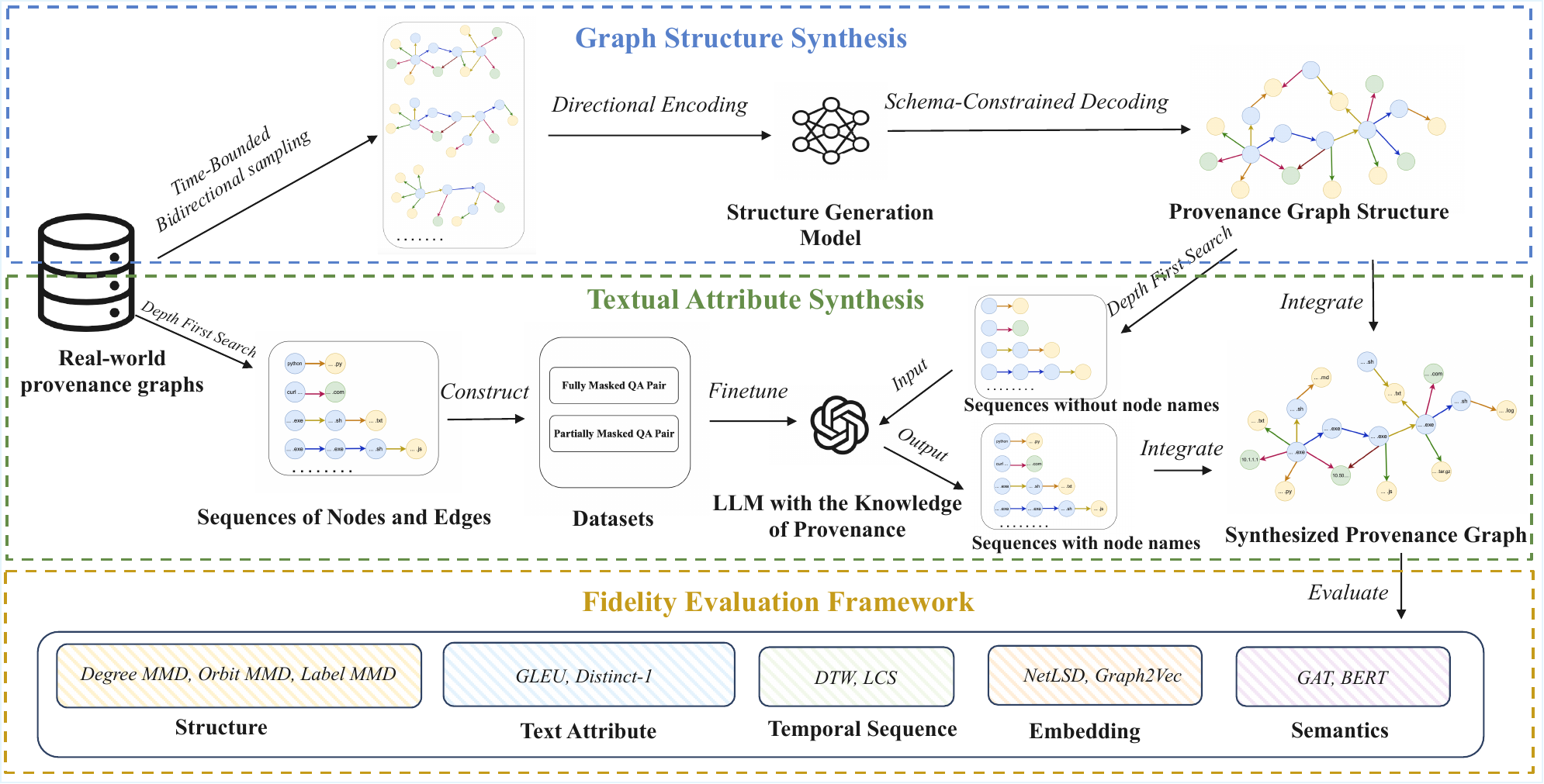} 
    \caption{\Sys Architecture. First, a heterogeneous graph generation model constructs the initial structure, which is subsequently refined through rule-based topological constraints. Next, an LLM is fine-tuned on serialized provenance data to capture domain-specific semantics, enabling it to populate the generated nodes with synthesized textual attributes. Finally, the resulting graphs are validated via a fidelity evaluation framework across five distinct dimensions. }
    \label{fig:overview} 
\end{figure*}

\section{Threat Model and Design Goal}
\PP{Threat Model.} We follow the same threat model as past \ac{pids} work~\cite{li2023nodlink,rehman2024flash,jia2024magic,jiang2025orthrus}. We assume attackers originate externally and aim to compromise internal systems for valuable data or persistent control. Their actions, though potentially complex, are assumed to leave observable traces in system logs monitored by kernel-level tools. The system hardware, OS, and core auditing software form our trusted base~\cite{custos,nitro}. 
Poison attacks and evasion attacks are not considered in our threat model~\cite{shoaib2026spectra}.

\PP{Design Goal.} \Sys is designed as a framework that integrates the complementary strengths of graph generation models for structural synthesis and large language models for text attribute generation. The objectives of \Sys are threefold: (1) to synthesize high-fidelity provenance graphs. High-fidelity provenance graphs require accurate modeling of both structural patterns and textual attributes;
(2) to mitigate data imbalance in existing provenance graph datasets. \Sys is designed to learn the underlying distribution of real-world data and generate samples that represent underrepresented scenarios, thereby improving data diversity; and (3) to enhance the performance of downstream \ac{pids}. By augmenting existing datasets with \Sys-generated graphs, the coverage of system behaviors can be expanded, leading to improved generalization and robustness of \ac{pids} models.

\section{\Sys Design}
In this section, we introduce the design details of \Sys. Figure~\ref{fig:overview} shows the workflow of \Sys, which consists of three main components: graph structure synthesis, textual attribute synthesis, and fidelity evaluation framework. In the first stage, we train a heterogeneous graph generation model using real provenance graph datasets to generate the graph structure. In the second stage, we fine-tune a large language model to label the nodes with appropriate text attributes. Finally, we evaluate the fidelity of the synthesized provenance graphs from five perspectives.

\subsection{Graph Structure Synthesis}
In the first stage, we focus on generating the structure of provenance graphs. We train an enhanced GraphGen model with a direction-aware DFS encoding to capture edge directions, and a schema-constrained decoder to ensure the validity of the generated structures. To make this model applicable to large provenance graphs, we apply a time-bounded bidirectional subgraph sampling strategy to extract representative subgraphs.

\PP{Structure Generation Model.}
Existing graph generation models primarily focus on structural connectivity, often overlooking the modeling of node and edge labels~\cite{de2018molgan,you2018graph}. Many approaches also depend on domain-specific heuristics (e.g., those tailored to molecular graphs), limiting their generalizability across graph types~\cite{samanta2020nevae}. In contrast, provenance graphs represent a new class of heterogeneous graphs, where nodes correspond to system entities and edges represent system call types. Generating such graphs requires a domain-agnostic model that captures both structural and label information.
To this end, we build our approach on GraphGen~\cite{goyal2020graphgen}, a domain-agnostic framework that generates graphs autoregressively by jointly modeling structure and labels. However, the original GraphGen is not directly suitable for provenance graphs. To bridge this gap, we introduce two key enhancements: direction-aware structure modeling and schema-consistent relation generation.

First, we propose a direction-aware DFS encoding for provenance graphs. 
GraphGen represents each edge using a DFS-based tuple designed for undirected graphs. This representation is inadequate for provenance graphs, where edge direction is essential to describe information flow and causal dependencies. To resolve this, we extend the original 5-tuple into a direction-aware 6-tuple by adding a direction indicator, $D_e$. In this representation, $t_u$ and $t_v$ denote the DFS timestamps of the nodes, $L_u$ and $L_v$ are their respective labels, and $L_e$ is the edge label. The resulting 6-tuple for an edge at step $t$ is defined as:
\[
s_t = (t_u, t_v, L_u, D_e, L_e, L_v).
\]

Second, we introduce schema-constrained decoding to ensure semantic validity. Unlike generic graphs, provenance graphs must follow strict schema rules: only specific edge labels are allowed between certain types of source and destination nodes. A standard decoder might generate "invalid" connections (e.g., a \texttt{process-clone-network} relation) that violate these rules. To address this, we define a valid label set $\mathcal{M}(L_u, L_v)$ for any pair of node labels $(L_u, L_v)$ based on the structural regularities observed in real-world provenance graphs (detailed schemas for DARPA E3/E5 and OpTC are provided in Appendix \ref{sec:schema}, Table \ref{tab:schema}). During generation, we use an indicator function $\mathbf{1}[\cdot]$ to mask the model's hidden state $h_t$, ensuring that the probability of choosing an edge label $L_e$ is zero if it does not conform to the schema. The adjusted probability $p'$ is calculated as:
\[
p'(L_e \mid h_t, L_u, L_v) \propto p(L_e \mid h_t) \cdot \mathbf{1}[L_e \in \mathcal{M}(L_u, L_v)].
\]

\begin{algorithm}[!t]
\scriptsize
\DontPrintSemicolon
\SetKwInOut{Input}{Input}
\SetKwInOut{Output}{Output}

\Input{Directed graph $G=(V,E)$, seed node $u$, seed time $\tau_u$, time windows $\Delta_{\text{in}}, \Delta_{\text{out}}$, iteration limit $T$, sampling budgets $b_{\text{in}}, b_{\text{out}}$, size constraints}
\Output{Sampled subgraph $G_{\text{sampled}}$ or $\emptyset$}

$V_{\text{sub}} \leftarrow \{u\},\; E_{\text{sub}} \leftarrow \emptyset,\; \mathit{Frontier} \leftarrow \{u\}$\;
$\tau_{\min} \leftarrow \tau_u - \Delta_{\text{in}},\; \tau_{\max} \leftarrow \tau_u + \Delta_{\text{out}}$\;

\For{$t \leftarrow 1$ \KwTo $T$}{
    $C_{\text{in}} \leftarrow$ valid incoming edges of $\mathit{Frontier}$ within $[\tau_{\min}, \tau_{\max}] \setminus E_{\text{sub}}$\;
    $C_{\text{out}} \leftarrow$ valid outgoing edges of $\mathit{Frontier}$ within $[\tau_{\min}, \tau_{\max}] \setminus E_{\text{sub}}$\;
    
    \If{$C_{\text{in}} = \emptyset$ \textbf{and} $C_{\text{out}} = \emptyset$}{
        \textbf{break}\;
    }

    $S \leftarrow \text{UniformSample}(C_{\text{in}}, b_{\text{in}})\; \cup\; \text{UniformSample}(C_{\text{out}}, b_{\text{out}})$\;
    Add sampled edges in $S$ and their incident nodes to $(V_{\text{sub}}, E_{\text{sub}})$\;
    $\mathit{Frontier} \leftarrow$ newly discovered nodes from $S$\;

    \If{$\mathit{Frontier} = \emptyset$ \textbf{or} $|V_{\text{sub}}| \geq \mathit{max\_nodes}$ \textbf{or} $|E_{\text{sub}}| \geq \mathit{max\_edges}$}{
        \textbf{break}\;
    }
}

Construct $G_{\text{sampled}}$ from $(V_{\text{sub}}, E_{\text{sub}})$, remove self-loops, and relabel nodes\;

\If{$G_{\text{sampled}}$ is not weakly connected or violates size constraints}{
    \Return $\emptyset$\;
}
\Return $G_{\text{sampled}}$\;

\caption{\sc{Time-Bounded Bidirectional Subgraph Sampling.}}
\label{alg:subgraph_sampling}
\end{algorithm}

\PP{Time-Bounded Bidirectional Subgraph Sampling.}
GraphGen is typically trained on small graphs with around hundreds of nodes, whereas provenance graphs often contain millions of nodes, making it infeasible to feed the entire graph into the model. To make training scalable on large provenance graphs, we adopt a subgraph sampling strategy and use the resulting samples to construct a GraphGen-compatible training dataset.

Unlike generic graph sampling methods such as node sampling or random walks, provenance subgraph sampling should preserve both \emph{temporal locality} and \emph{directed dependency structure}. In provenance graphs, semantically related activities are typically concentrated within a limited time span and connected through directed interactions among system entities. To capture such patterns, we propose a \emph{time-bounded bidirectional subgraph sampling} strategy. Given a seed node \(u\) with timestamp \(\tau_u\), we first restrict the candidate search space to edges whose timestamps fall within a temporal window \([\tau_u - \Delta_{\text{in}}, \tau_u + \Delta_{\text{out}}]\), where \(\Delta_{\text{in}}\) and \(\Delta_{\text{out}}\) denote the backward and forward time horizons, respectively. We then iteratively expand the sampled subgraph from the current frontier by selecting both incoming and outgoing edges within this temporal window, so that both upstream dependencies and downstream effects of the seed can be incorporated into the sample. In this way, the resulting subgraph remains temporally focused while preserving the dependency context surrounding the seed node.
The complete sampling procedure is presented in Algorithm~\ref{alg:subgraph_sampling}.

\subsection{Textual Attribute Synthesis}
After the first stage, we obtain a directed graph in which nodes of different types are connected by edges of different types, and all connections conform to predefined rules. In this stage, we focus on assigning attributes to the nodes in the generated graph to synthesize complete provenance graphs.

We leverage the natural language generation capabilities of large language models to assign names to nodes in provenance graphs. This process involves two main challenges. First, prior studies suggest that LLMs are not inherently effective at understanding graph structures ~\cite{wang2023can,guo2023gpt4graph,jin2024large}. Thus, a key issue is how to transform the structure of a provenance graph into a format that is more interpretable by LLMs. Second, LLMs lack prior knowledge of typical node names in provenance graphs, and prompting them directly often results in overly simplistic outputs with limited diversity.

To address the first challenge, prior work has shown that LLMs are less effective at processing graph-structured data but perform well on sequence-based inputs~\cite{ye2023language,zhao2023graphtext,fatemi2023talk}. Motivated by this, we propose a method that converts directed heterogeneous graphs into sequences of node and edge types using depth-first search. These sequences are then input to the LLM to generate names for the corresponding nodes. Specifically, to extract complete and non-overlapping DFS paths, we use all nodes with zero in-degree as starting points for DFS traversal. 
In each extracted sequence, a node is represented as a pair $(\tau_i, n_i)$, where $\tau_i \in \mathcal{T}$ denotes the node type and $n_i \in \mathcal{N}$ the node name. Each edge $e_j$ represents the interaction type between adjacent nodes. A DFS-derived sequence is defined as:

$$
S = [(\tau_1, n_1) \xrightarrow{e_1} (\tau_2, n_2) \xrightarrow{e_2} \cdots \xrightarrow{e_{k-1}} (\tau_k, n_k)]
$$

For nodes without names, we use [null] as a placeholder. During the name generation process, we do not input all sequences to the LLM in parallel. This is because nodes may appear in multiple sequences, and parallel processing could lead to inconsistencies in assigned names. Instead, we sequentially feed each DFS-derived sequence to the LLM. Once a node name is generated, it is filled into the graph, ensuring that subsequent traversals can retrieve the updated name.

To address the second challenge, we construct a training dataset to provide the LLM with prior knowledge about entity names in provenance data. Each training instance is a question–answer pair, where the question (Q) consists of a DFS sequence of node and edge types, and the answer (A) is the same sequence with node names annotated. All training instances are derived from real-world provenance graphs.

To generate realistic Q\&A pairs, we adopt a masking strategy that preserves structural context while requiring the model to infer node names. 
During the inference process, two types of sequence are provided to the LLM. The \textit{first} type consists of sequences in which all nodes are unnamed, serving as a cold start and requiring the LLM to infer node names based solely on the order of node and edge types. The \textit{second} type includes sequences where some nodes are already named, requiring the LLM to infer the remaining names using both structural order and existing node names as context. Accordingly, we define two corresponding Q\&A pair formats:

\textbf{Fully Masked Q\&A Pairs}
In this setting, all node names are masked with [null] while retaining node types and edge types, creating a template for cold-start name generation. This design encourages the model to learn a mapping $f: (\tau_i, \{\tau_j, e_l\}_{j=1}^k) \rightarrow n_i$, while preserving structural consistency through fixed sequences of node and edge types.
$$
\begin{aligned}
Q_{\text{full}} &= [(\tau_1, \text{[null]}) \xrightarrow{e_1} (\tau_2, \text{[null]}) \xrightarrow{e_2} \cdots \xrightarrow{e_{k-1}} (\tau_k, \text{[null]})], \\
A_{\text{full}} &= [(\tau_1, n_1) \xrightarrow{e_1} (\tau_2, n_2) \xrightarrow{e_2} \cdots \xrightarrow{e_{k-1}} (\tau_k, n_k)].
\end{aligned}
$$

\textbf{Partially Masked Q\&A Pairs}
To simulate realistic generation scenarios where some node names are known from prior context, we randomly mask a subset $\mathcal{M} \subset \{1,\ldots,k\}$ of node names with a masking rate $\rho$:

$$
Q_{\text{part}} = [(\tau_1, \tilde{n}_1) \xrightarrow{e_1} \cdots \xrightarrow{e_{k-1}} (\tau_k, \tilde{n}_k)],
$$

where

$$
\tilde{n}_i = \begin{cases} 
\text{[null]} & \text{if } i \in \mathcal{M}, \\
n_i & \text{otherwise}.
\end{cases}
$$

The corresponding answer sequence $A_{\text{part}} = A_{\text{full}}$ provides full supervision of node names. This adaptive masking strategy allows the model to learn to predict masked node names based on both the observed names and the structural context:

$$
P(n_{\mathcal{M}} \mid n_{\neg\mathcal{M}}, \{\tau_j\}_{j=1}^k, \{e_l\}_{l=1}^{k-1}).
$$

By constructing these two types of Q\&A  pairs, we provide the LLM with a comprehensive training dataset that helps it learn the correspondence between sequences and node names, thereby improving its ability to generate accurate node names in provenance graphs. This approach leverages the strengths of LLMs in sequence-based tasks while addressing the challenges of graph structure understanding, leading to more effective node name generation in provenance graphs.

\subsection{Fidelity Evaluation Framework}
\label{fidelity}
To comprehensively assess the fidelity of synthetic provenance graphs, we propose a framework that evaluates synthesized graphs from complementary perspectives, including structure, text attribute, temporal sequence, embedding, and semantics. Collectively, these evaluation protocols quantify the extent to which synthetic samples can serve as reliable surrogates for empirical system activities.

\PP{Structural Evaluation.} To quantitatively evaluate the structural fidelity of generated graphs, we employ Maximum Mean Discrepancy (MMD)-based metrics~\cite{gretton2006kernel,thompson2022evaluation} that measure distributional divergence between generated graphs and real graphs. These metrics include: (1) \textit{degree MMD} for local connectivity patterns, (2) \textit{orbit MMD} for higher-order structural motifs, (3) \textit{edge label MMD} for edge label distributions, and (4) \textit{node label–degree MMD} for label–topology interactions.
The MMD² values are computed via the unbiased estimator, where \(P\) and \(Q\) denote reference and generated graph distributions, \(x,x'\) are samples from \(P\), and \(y,y'\) from \(Q\), and $k(\cdot, \cdot)$ is a kernel function: 

\begin{align*}
\text{MMD}^2 =\ & \mathbb{E}_{x,x'\sim P}[k(x,x')] + \mathbb{E}_{y,y'\sim Q}[k(y,y')] \\
& - 2\mathbb{E}_{x\sim P, y\sim Q}[k(x,y)]
\end{align*}

\PP{Textual Evaluation.} To evaluate the fidelity and diversity of generated text attributes, we adopt two standard NLP metrics: (1) \textit{GLEU}~\cite{mutton2007gleu}, which evaluates fidelity through n-gram overlap and is effective for short sequences; and (2) \textit{Distinct-1}~\cite{li2016diversity}, which measures lexical diversity by computing the proportion of unique unigrams in the generated text.
To adapt GLEU for graph textual attributes, we employ a reference-based greedy matching approach~\cite{rus2012optimal,kusner2015word,zhang2019bertscore} that focuses on the highest match to avoid dilution from irrelevant references. For each node type $t$, we construct a reference corpus $\mathcal{R}_t = \{r_1, r_2, \ldots, r_n\}$ by aggregating all valid node names from the real-world provenance graph. For a set of generated names $\mathcal{G}_t = \{g_1, g_2, \ldots, g_{|\mathcal{G}_t|}\}$, we compute similarities between each $g_i$ and all $r_j \in \mathcal{R}_t$, select the maximum similarity via greedy matching, and average these values:
\[
S_t = \frac{1}{|\mathcal{G}_t|} \sum_{g_i \in \mathcal{G}_t}
\max_{r_j \in \mathcal{R}_t} \text{GLEU}(g_i, r_j).
\]

\PP{Temporal Evaluation.} A provenance graph is a directed graph in which directed sequences also represent temporal sequences of events occurring in chronological order. For synthetic provenance graphs, accurately preserving the temporal structure of event sequences is essential for ensuring fidelity. To evaluate the temporal consistency of synthetic provenance graphs, we employ two metrics: \textit{Longest Common Subsequence (LCS)}~\cite{bergroth2000survey} and \textit{Dynamic Time Warping (DTW)}~\cite{muller2007dynamic}.
LCS measures the longest common event sequence through exact and strict alignment, while DTW offers flexibility by allowing non-linear, stretched, or compressed sequence alignments along the time axis.

\PP{Embedding Evaluation.} To evaluate the fidelity of synthetic graphs from an embedding perspective, we employ graph representation learning techniques to project both synthetic and real graphs into a shared vector space, and measure their cosine similarity. Specifically, we adopt two representative embedding methods: \textit{NetLSD}~\cite{tsitsulin2018netlsd} and \textit{Graph2Vec}~\cite{narayanan2017graph2vec}. To assess the similarity between a synthetic graph $G_s$ and a real graph $G_r$, we compute the cosine similarity between their respective embeddings:
$$
\text{Similarity} = \frac{\mathbf{v}_{G_s} \cdot \mathbf{v}_{G_r}}{\|\mathbf{v}_{G_s}\| \|\mathbf{v}_{G_r}\|}
$$

\PP{Semantic Evaluation.}
\label{sec:semantic}
Evaluating semantic correctness is essential for assessing the fidelity of synthetic provenance graphs, as it indicates whether the generated graphs accurately reflect the underlying logic and intent of real-world system behaviors. Despite its importance, this evaluation remains difficult due to the lack of standardized metrics and the inherent complexity in defining ground-truth semantics for diverse and dynamic system activities.

To evaluate semantic correctness, we formally define a semantic unit as a triplet \((s, e, t)\), where \(s, t \in \mathcal{N}\) are node names and \(e \in \mathcal{E}\) is the edge type. Concatenating these elements forms a natural language sentence \(\text{Sentence}(s, e, t)\), capturing the semantics of the interaction between entities. We employ a BERT~\cite{devlin2019bertpretrainingdeepbidirectional}-based encoder to obtain its embedding, formally defined as:
\[
\mathbf{v} = \text{BERT}(\text{Sentence}(s, e, t))
\]
Since this embedding captures only pairwise relationships, we further employ a multi-layer Graph Attention Network (GAT)~\cite{velickovic2017graph} to achieve graph-level assessment. Each node $v$ with name $T_v$ is initialized as $\mathbf{h}_v^{(0)} = \text{BERT}(T_v)$, and the GAT aggregates neighborhood information through stacked self-attention layers. For a semantic unit \((s, e, t)\), we obtain global embeddings:
\[
\mathbf{u}_s = \text{GAT}\left(s; \left\{ \mathbf{h}_v^{(0)} \right\}_{v\in\mathcal{V}}\right), \quad \mathbf{u}_t = \text{GAT}\left(t; \left\{ \mathbf{h}_v^{(0)} \right\}_{v\in\mathcal{V}}\right)
\]
These are concatenated with the local embedding to form a comprehensive representation $\mathbf{z} = [\mathbf{v}; \mathbf{u}_s; \mathbf{u}_t]$ capturing both local and global context. To quantify semantic correctness, we construct a reference corpus $\mathcal{Z}_{\text{ref}} = \{\mathbf{z}_1, \ldots, \mathbf{z}_m\}$ from real-world provenance graphs. Using the same greedy matching strategy~\cite{rus2012optimal,kusner2015word,zhang2019bertscore} as in textual evaluation, the semantic correctness score is computed as:
\[
S_{\text{sem}} = \frac{1}{|\mathcal{Z}_{\text{gen}}|} \sum_{\mathbf{z} \in \mathcal{Z}_{\text{gen}}}
\max_{\mathbf{z}_j \in \mathcal{Z}_{\text{ref}}} \text{sim}(\mathbf{z}, \mathbf{z}_j).
\]

\section{Evaluation}
To thoroughly evaluate \Sys, we address the following research questions. Experiments were performed on a machine with Intel Xeon Platinum CPU, 512 GB RAM, NVIDIA RTX A6000 GPUs, and Ubuntu 22.04.3 LTS. The research questions are:

\textbf{RQ1:} What is the fidelity of synthetic provenance graphs in terms of structural, textual, temporal, embedding, and semantic characteristics?

\textbf{RQ2:} Can synthetic provenance graphs mitigate class imbalance in the dataset?

\textbf{RQ3:} When synthetic provenance graphs are used for training set augmentation in downstream intrusion detection tasks, what is the performance of the detection task?

\textbf{RQ4:} What are the time and resource costs of generating synthetic provenance graphs?

Further aspects of our study, including hyperparameter analysis and ablation studies, are discussed in Appendix~\ref{sec:hyper} and ~\ref{sec:ablation}.

\subsection{Experimental Setup}
\PP{Datasets.}
We evaluate the effectiveness of \Sys on three public datasets: the DARPA Engagement 3 dataset~\cite{darpa-E3}, the DARPA Engagement 5 dataset~\cite{darpa-E5}, and the OpTC dataset~\cite{OpTC-data}, utilizing the labels from Orthrus~\cite{jiang2025orthrus}. From each of these datasets, we select two subsets, resulting in a total of six datasets: Cadets-E3, Theia-E3, ClearScope-E5, Theia-E5, OpTC-H501, and OpTC-H201. The provenance graphs in these datasets contain three types of nodes and ten types of events, with the number of nodes ranging from 50\,K to 2,000\,K and the number of edges ranging from 100\,K to 8,000\,K. For each dataset, we use provenance data from the first day for training and fidelity evaluation, with the two sets being derived from distinct time windows and strictly separated.

\PP{Baselines.}
In our approach, we first use a heterogeneous graph generation model to construct the structure of the provenance graph and then employ an LLM to generate the textual attributes of the graph. In contrast, an alternative method directly calls an LLM to synthesize provenance logs and then converts these logs into a graph. This method does not require an additional heterogeneous graph generation model. We systematically evaluate and compare the provenance data synthesized by LLMs with the data synthesized by \Sys. 

To ensure a competitive comparison, we selected a set of strong baseline models, including GPT-5.1~\cite{openai-gpt51}, Claude-sonnet-4.5~\cite{claude}, DeepSeek-v3.2~\cite{liu2025deepseek}, and Qwen3-max~\cite{yang2025qwen3}. These models are widely considered state-of-the-art LLMs at the time of evaluation. To generate provenance logs using the LLMs listed above, we designed structured prompts that included task descriptions and illustrative examples. Each dataset was paired with a customized prompt, primarily differing in the examples provided. These examples contained five elements: source node type, source node name, destination node type, destination node name, and edge type. They were selected from real audit logs to ensure both representativeness and coverage of various node and edge types. For each dataset, we collected 100 responses from each LLM. After obtaining the responses, we filtered out entries that contain node or edge types outside of the predefined type sets. The remaining entries were then used to construct graphs for comparison with graphs generated by \Sys. The prompt used in our experiments is provided in Appendix~\ref{sec:prompt} (Figure~\ref{fig:prompt}).

\begin{table*}[t!]
\centering
\caption{Comprehensive Comparison of MMD, Temporal, and Embedding Metrics. For MMD and DTW, lower is better. For LCS, NetLSD, and Graph2Vec, higher is better. Bold values indicate the best performance in each category.}
\resizebox{\textwidth}{!}{%
\begin{tabular}{llcccccccccc}
\toprule
\multirow{2}{*}{\textbf{Dataset}} & \multirow{2}{*}{\textbf{Model}} & \multicolumn{4}{c}{\textbf{MMD Metrics ($\downarrow$)}} & \multicolumn{2}{c}{\textbf{Temporal Metrics}} & \multicolumn{2}{c}{\textbf{Embedding Metrics ($\uparrow$)}} \\
\cmidrule(lr){3-6} \cmidrule(lr){7-8} \cmidrule(lr){9-10}
& & \textbf{Degree} & \textbf{Orbit} & \textbf{Edge L.} & \textbf{Node L.+Deg} & \textbf{LCS ($\uparrow$)} & \textbf{DTW ($\downarrow$)} & \textbf{NetLSD} & \textbf{Graph2Vec} \\
\midrule
\multirow{5}{*}{\textbf{Cadets-E3}} 
& \textbf{GPT-5.1} & 0.59 & 0.87 & 0.41 & 1.16 & 3.02 & 1.58 & 0.48 & 0.29 \\
& \textbf{Claude-sonnet-4.5} & 1.01 & 0.96 & 0.35 & 1.28 & 4.07 & 1.62 & 0.16 & 0.33 \\
& \textbf{DeepSeek-v3.2} & 1.47 & 0.88 & 0.72 & 1.45 & 3.03 & 1.69 & 0.03 & 0.34 \\
& \textbf{Qwen3-max} & 1.36 & 0.94 & 0.10 & 1.28 & 3.00 & \textbf{0.00} & 0.03 & 0.29 \\
& \textbf{ProvSyn} & \textbf{0.37} & \textbf{0.79} & \textbf{0.01} & \textbf{0.61} & \textbf{4.71} & 0.74 & \textbf{0.49} & \textbf{0.35} \\
\midrule
\multirow{5}{*}{\textbf{Theia-E3}} 
& \textbf{GPT-5.1} & 0.28 & 0.37 & 0.17 & 0.60 & 3.14 & 0.53 & 0.34 & 0.36 \\
& \textbf{Claude-sonnet-4.5} & 0.42 & 0.33 & 0.19 & 0.73 & 3.00 & \textbf{0.00} & 0.83 & 0.34 \\
& \textbf{DeepSeek-v3.2} & 0.57 & 0.38 & 0.29 & 0.65 & 2.93 & 1.92 & 0.42 & 0.38 \\
& \textbf{Qwen3-max} & 0.81 & 0.90 & 0.22 & 1.07 & 3.00 & \textbf{0.00} & \textbf{0.85} & 0.31 \\
& \textbf{ProvSyn} & \textbf{0.07} & \textbf{0.19} & \textbf{0.07} & \textbf{0.29} & \textbf{3.19} & 0.94 & 0.53 & \textbf{0.47} \\
\midrule
\multirow{5}{*}{\textbf{ClearScope-E5}} 
& \textbf{GPT-5.1} & 0.73 & 0.47 & 0.01 & 1.16 & 3.00 & 1.64 & 0.54 & 0.18 \\
& \textbf{Claude-sonnet-4.5} & 0.61 & 1.01 & 0.02 & 1.11 & 3.00 & 1.71 & 0.60 & 0.19 \\
& \textbf{DeepSeek-v3.2} & 0.56 & 0.49 & 0.01 & 0.93 & 3.00 & 1.64 & 0.41 & 0.19 \\
& \textbf{Qwen3-max} & 0.57 & 0.59 & \textbf{0.00} & 0.64 & 3.00 & 1.66 & 0.62 & 0.19 \\
& \textbf{ProvSyn} & \textbf{0.22} & \textbf{0.38} & \textbf{0.00} & \textbf{0.23} & \textbf{3.23} & \textbf{1.48} & \textbf{0.73} & \textbf{0.34} \\
\midrule
\multirow{5}{*}{\textbf{Theia-E5}} 
& \textbf{GPT-5.1} & 0.66 & 0.60 & 0.12 & 0.67 & 3.03 & 0.45 & 0.42 & 0.37 \\
& \textbf{Claude-sonnet-4.5} & 0.60 & 0.72 & 0.08 & 0.81 & \textbf{3.21} & 0.46 & 0.36 & 0.37 \\
& \textbf{DeepSeek-v3.2} & 0.70 & 1.10 & 0.43 & 0.54 & 3.02 & 0.53 & 0.10 & \textbf{0.40} \\
& \textbf{Qwen3-max} & 0.66 & 1.06 & 0.18 & 0.69 & 3.03 & 0.42 & 0.23 & 0.37 \\
& \textbf{ProvSyn} & \textbf{0.13} & \textbf{0.29} & \textbf{0.03} & \textbf{0.20} & 3.02 & \textbf{0.35} & \textbf{0.53} & 0.35 \\
\midrule
\multirow{5}{*}{\textbf{OpTC-H201}} 
& \textbf{GPT-5.1} & 0.65 & 0.75 & 0.51 & 0.62 & 4.09 & 2.98 & 0.28 & 0.23 \\
& \textbf{Claude-sonnet-4.5} & 0.98 & 1.22 & 1.93 & 1.77 & 3.50 & 3.04 & 0.04 & 0.21 \\
& \textbf{DeepSeek-v3.2} & 0.59 & 0.96 & 1.17 & 1.15 & 2.86 & 3.32 & 0.01 & 0.28 \\
& \textbf{Qwen3-max} & 0.55 & 0.79 & 1.00 & 1.05 & 2.76 & 3.36 & 0.01 & 0.29 \\
& \textbf{ProvSyn} & \textbf{0.10} & \textbf{0.51} & \textbf{0.11} & \textbf{0.17} & \textbf{4.74} & \textbf{1.73} & \textbf{0.64} & \textbf{0.33} \\
\midrule
\multirow{5}{*}{\textbf{OpTC-H501}} 
& \textbf{GPT-5.1} & 0.85 & 1.08 & 1.13 & 1.04 & 3.86 & 2.56 & 0.14 & 0.31 \\
& \textbf{Claude-sonnet-4.5} & 0.45 & 1.09 & 0.92 & 1.17 & 3.90 & 2.74 & 0.03 & 0.26 \\
& \textbf{DeepSeek-v3.2} & 0.77 & 1.19 & 0.93 & 1.52 & 2.45 & 3.71 & 0.01 & 0.37 \\
& \textbf{Qwen3-max} & 0.67 & 1.09 & 1.54 & 1.38 & 2.64 & 3.30 & 0.01 & 0.37 \\
& \textbf{ProvSyn} & \textbf{0.08} & \textbf{0.52} & \textbf{0.02} & \textbf{0.11} & \textbf{4.65} & \textbf{1.78} & \textbf{0.61} & \textbf{0.40} \\
\bottomrule
\end{tabular}
}
\label{tab:mmd_temporal_embedding}
\end{table*}

\PP{Metrics.}
For fidelity evaluation, we adopt the five-dimensional metrics introduced in Section~\ref{fidelity}.
For class imbalance mitigation, we use three metrics: \textit{novelty}, \textit{uniqueness}, and \textit{normalized entropy}. Novelty verifies that each generated graph is neither a subgraph of any training reference graph nor contains any reference graph as a subgraph, indicating the presence of novel structures. Uniqueness measures the proportion of generated graphs that are structurally distinct from one another, reflecting diversity. Normalized entropy measures label uncertainty and rescales it by the maximum entropy under the same number of classes, yielding values in the range $[0,1]$. Higher values indicate a more balanced label distribution.
For IDS application, beyond the standard F-score, we report three robust metrics for comprehensive evaluation: \textit{Balanced Accuracy (B.Acc.)}, which reduces the impact of imbalanced test data; the \textit{Matthews Correlation Coefficient} (MCC)~\cite{jiang2025orthrus}, which measures the correlation between predicted and true labels; and the \textit{Diagnostic Odds Ratio (DOR)}, which reflects diagnostic performance as the odds ratio of positive results for attacks versus normal behaviors.

\PP{Configurations.}
For training the graph generation model, we sampled 20,000 subgraphs from the real provenance graphs. Among them, 16,000 subgraphs were used for training, 2,000 for validation, and 2,000 for testing. The model was trained for 3,000 epochs with a batch size of 32 and a learning rate of 0.003. For the text attribute synthesis task, we used Llama3.2-3B-Instruct~\cite{grattafiori2024llama} as the base model. We constructed a training set of 20,000 samples, consisting of 10,000 fully masked Q\&A  pairs and 10,000 partially masked Q\&A  pairs. During training, we applied the LoRA fine-tuning method, freezing most of the model parameters and updating only 0.81\% of them. The model was trained for 60 steps with a learning rate of 2e-4. During inference, the temperature was set to 1.5, and the maximum number of generated tokens was set to 2048.

\subsection{Fidelity Evaluation}
\label{subsubsec:structure-eval}

\begin{table*}[t]
\centering
\small
\setlength{\tabcolsep}{2.5pt} 
\renewcommand{\arraystretch}{0.9} 
\caption{Comparison of textual quality across different models, node types, and datasets. Left: GLEU. Right: Distinct-1. The best value in each group is bolded. Here, GPT, Claude, DeepSeek, and Qwen3 serve as abbreviations for the specific model versions detailed in Table~\ref{tab:mmd_temporal_embedding}.}
\resizebox{\textwidth}{!}{%
\begin{tabular}{llccccc|ccccc}
\toprule
\multirow{2}{*}{\textbf{Dataset}} & \multirow{2}{*}{\textbf{Type}} 
& \multicolumn{5}{c}{\textbf{GLEU}} 
& \multicolumn{5}{c}{\textbf{Distinct-1}} \\
\cmidrule(lr){3-7} \cmidrule(lr){8-12}
 &  
 & \textbf{GPT} & \textbf{Claude} & \textbf{DeepSeek} & \textbf{Qwen3} & \textbf{ProvSyn}
 & \textbf{GPT} & \textbf{Claude} & \textbf{DeepSeek} & \textbf{Qwen3} & \textbf{ProvSyn} \\
\midrule
\multirow{2}{*}{\textbf{Cadets-E3}} 
 & \textbf{Process} & 0.36 & 0.16 & 0.04 & 0.07 & \textbf{0.56}
 & 1.00 & 1.00 & 1.00 & 1.00 & \textbf{1.00} \\
 & \textbf{File} & 0.10 & 0.07 & 0.02 & 0.03 & \textbf{0.22}
 & 1.00 & 1.00 & 1.00 & 1.00 & \textbf{1.00} \\
\midrule
\multirow{2}{*}{\textbf{Theia-E3}} 
 & \textbf{Process} & 0.03 & 0.14 & 0.00 & 0.50 & \textbf{0.69}
 & 0.66 & 1.00 & 0.51 & \textbf{1.00} & 0.70 \\
 & \textbf{File} & 0.01 & 0.00 & 0.00 & 0.00 & \textbf{0.06}
 & 1.00 & 1.00 & 1.00 & 1.00 & \textbf{1.00} \\
\midrule
\multirow{2}{*}{\textbf{ClearScope-E5}} 
 & \textbf{Process} & 0.47 & \textbf{0.78} & 0.37 & 0.33 & 0.52
 & 0.55 & 0.67 & 0.51 & 0.57 & \textbf{0.68} \\
 & \textbf{File} & 0.08 & 0.06 & 0.06 & 0.08 & \textbf{0.21}
 & 0.75 & 0.81 & 0.73 & 0.73 & \textbf{0.84} \\
\midrule
\multirow{2}{*}{\textbf{Theia-E5}} 
 & \textbf{Process} & 0.04 & 0.05 & 0.03 & 0.05 & \textbf{0.18}
 & 0.54 & 0.60 & 0.38 & 0.57 & \textbf{0.83} \\
 & \textbf{File} & 0.03 & 0.03 & 0.03 & 0.03 & \textbf{0.03}
 & 1.00 & 1.00 & 1.00 & 1.00 & \textbf{1.00} \\
\midrule
\multirow{2}{*}{\textbf{OpTC-H201}} 
 & \textbf{Process} & 0.21 & 0.33 & 0.19 & 0.19 & \textbf{0.55}
 & 0.31 & 0.38 & 0.22 & 0.40 & \textbf{0.46} \\
 & \textbf{File} & 0.17 & 0.15 & 0.15 & 0.19 & \textbf{0.30}
 & 0.70 & 0.94 & 0.95 & \textbf{0.99} & 0.71 \\
\midrule
\multirow{2}{*}{\textbf{OpTC-H501}} 
 & \textbf{Process} & 0.33 & 0.36 & 0.23 & 0.31 & \textbf{0.38}
 & 0.20 & 0.25 & 0.34 & 0.31 & \textbf{0.39} \\
 & \textbf{File} & 0.00 & 0.00 & 0.00 & 0.00 & 0.00
 & 0.69 & 0.76 & 0.80 & 0.79 & \textbf{0.84} \\
\bottomrule
\end{tabular}%
}
\label{tab:textual}
\end{table*}

\PP{Structural Evaluation.}
Table \ref{tab:mmd_temporal_embedding} shows structural fidelity via MMD metrics. \Sys consistently achieves the lowest scores across all datasets and distributions (degree, orbit, edge/node labels), indicating superior alignment with real graph structures. Notably, on OpTC-H201 and OpTC-H501, \Sys reduces MMD scores by over an order of magnitude compared to baselines. While baseline models exhibit higher and less stable scores, \Sys’s performance underscores the necessity of a dedicated heterogeneous graph generation network for faithful structural synthesis beyond standalone LLM capabilities.

\PP{Textual Evaluation.} 
We evaluated the textual quality of process and file nodes using GLEU and Distinct-1; network nodes were excluded as their IPv4-compliant numerical names are unsuitable for NLP metrics. As shown in Table \ref{tab:textual}, baselines struggle with OS entity naming (GLEU < 0.20), revealing a lack of domain-specific knowledge. Conversely, fine-tuned on real entity names, \Sys generates more accurate and realistic names. \Sys also outperforms baselines in Distinct-1 diversity, particularly in ClearScope-E5 and Theia-E5 where baselines produce monotonous names. This demonstrates \Sys's superior ability to capture the complex naming distributions of real-world operating systems.

\PP{Temporal Evaluation.} 
In our experiments, we used depth-first search to extract sequences from the graphs. Each sequence consisted of node types and edge types, e.g. \texttt{(process, clone, process, read, file)}. 
Table \ref{tab:mmd_temporal_embedding} shows that across six datasets, \Sys achieves higher LCS scores and lower DTW values than the baselines in five datasets. In particular, for Cadets-E3, OpTC-H201, and OpTC-H501, \Sys attains LCS scores above 4.5,  highlighting its effectiveness in maintaining temporal fidelity.

\PP{Embedding Evaluation.} 
As shown in Table~\ref{tab:mmd_temporal_embedding}, across six datasets, \Sys consistently achieves the highest relative similarity scores in five datasets for both NetLSD and Graph2Vec. Notably, on the OpTC datasets, where all baselines have NetLSD scores below 0.30, \Sys maintains scores above 0.60, demonstrating its robustness in preserving embedding fidelity.

\begin{figure*}[htbp] 
    \centering
    \includegraphics[width=\textwidth]{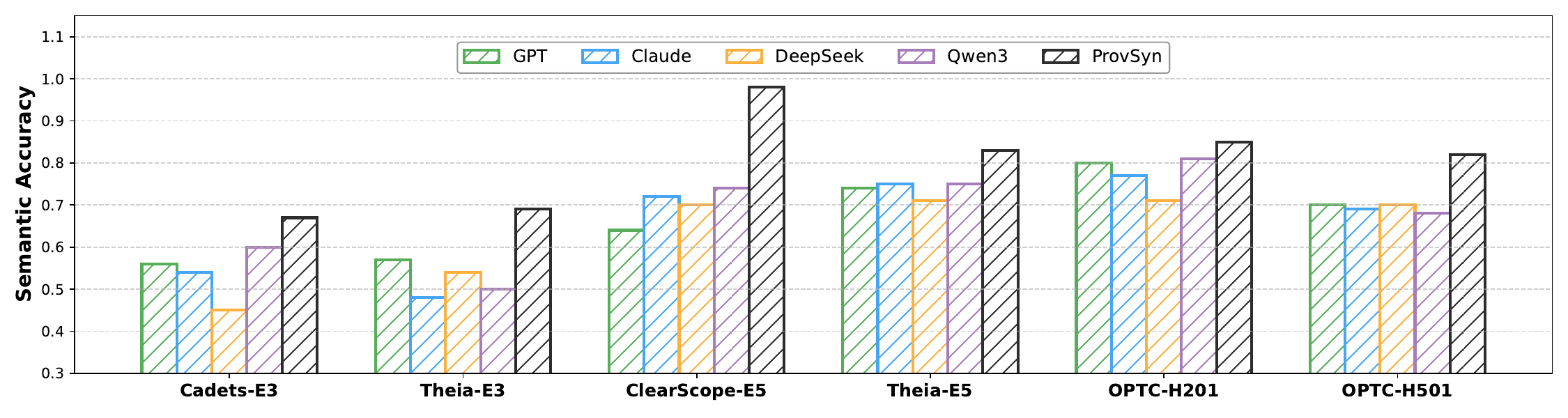}
    \caption{Comparison of semantic accuracy across different models and datasets. Higher values indicate better performance.} 
    \label{fig:semantic} 
\end{figure*}

\PP{Semantic Evaluation.}
To avoid dimensional bias, we project BERT node embeddings into the GAT output space via a linear layer before concatenation. As shown in Figure~\ref{fig:semantic}, \Sys consistently outperforms baselines in semantic accuracy across all six datasets, demonstrating robustness to distribution variations. Notably, \Sys achieves a substantial lead in datasets like ClearScope-E5, reaching 0.98 accuracy, significantly higher than the baselines.

\subsection{Data Imbalance Mitigation}
\begin{table}[t]
\centering
\caption{Evaluation of data imbalance mitigation. The left section includes two metrics: Novelty (Nov.) and Uniqueness (Uniq.). The middle and right sections compare Node/Edge Normalized Entropy (Real vs. Real + Synthetic), where higher values indicate better balance.}
\label{tab:imbalance_mitigation}
\footnotesize 
\setlength{\tabcolsep}{3pt} 
\begin{tabular*}{\columnwidth}{@{\extracolsep{\fill}} l cc cc cc}
\toprule
\multirow{2}{*}{\textbf{Dataset}} & \multicolumn{2}{c}{\textbf{Nov. \& Uniq.}} & \multicolumn{2}{c}{\textbf{Node Entropy}} & \multicolumn{2}{c}{\textbf{Edge Entropy}} \\
\cmidrule{2-3} \cmidrule{4-5} \cmidrule{6-7}
& \textbf{Nov.} & \textbf{Uniq.} & \textbf{Real} & \textbf{+\Sys} & \textbf{Real} & \textbf{+\Sys} \\
\midrule
\textbf{Cadets-E3}     & \textbf{100.0} & \textbf{100.0} & 0.34 & \textbf{0.64} & 0.40 & \textbf{0.46} \\
\textbf{Theia-E3}      & \textbf{100.0} & \textbf{100.0} & 0.91 & \textbf{0.92} & 0.59 & \textbf{0.59} \\
\textbf{ClearScope-E5} & \textbf{100.0} & \textbf{100.0} & 0.30 & \textbf{0.41} & 0.33 & \textbf{0.34} \\
\textbf{Theia-E5}      & \textbf{100.0} & \textbf{100.0} & 0.24 & \textbf{0.59} & 0.35 & \textbf{0.47} \\
\textbf{OpTC-H201}     & \textbf{100.0} & \textbf{100.0} & 0.43 & \textbf{0.70} & 0.56 & \textbf{0.57} \\
\textbf{OpTC-H501}     & \textbf{100.0} & \textbf{100.0} & 0.41 & \textbf{0.58} & 0.55 & 0.55          \\
\bottomrule
\end{tabular*}
\end{table}
\PP{Novelty and Uniqueness.} As shown in Table~\ref{tab:imbalance_mitigation}, the graphs generated by \Sys achieve 100\% novelty and 100\% uniqueness across all six datasets. This indicates that the synthetic data exhibits strong diversity and effectively captures patterns that were underrepresented or missing in the original data.

\PP{Normalized Entropy.} 
We compute normalized entropy for both the original datasets and the augmented datasets after incorporating \Sys-generated graphs. The results are summarized in Table~\ref{tab:imbalance_mitigation}. Overall, the inclusion of \Sys-generated data consistently increases the normalized entropy of node and edge label distributions across all datasets.
Notably, datasets with more severe initial imbalance exhibit more pronounced entropy improvements after augmentation. For example, in Theia-E5, the node label normalized entropy increases substantially from 0.24 to 0.59. These results suggest that \Sys contributes to a more balanced label distribution by enriching underrepresented classes.

\subsection{Application in IDS}
\begin{table*}[t]
\centering
\caption{Performance of IDS models trained on ``Real'' vs. ``+ ProvSyn'' (real + synthesized) data. Green/Red cells indicate performance improvement/decrease over the baseline. MCC is N/A for single-class predictions; DOR is $\infty$ for zero false positives or negatives. F-Score and MCC are scaled by $10^3$.}
\label{tab:ids}
\resizebox{0.9\textwidth}{!}{%
\begin{tabular}{llccccccccc}
\toprule
 & & \multicolumn{4}{c}{\textbf{NodLink}} & & \multicolumn{4}{c}{\textbf{Flash}} \\
\cmidrule(lr){3-6} \cmidrule(lr){8-11}
\textbf{Dataset} & \textbf{Source} & \textbf{F-Score} & \textbf{B.Acc.} & \textbf{MCC} & \textbf{DOR} & & \textbf{F-Score} & \textbf{B.Acc.} & \textbf{MCC} & \textbf{DOR} \\
\midrule
\multirow{2}{*}{\textbf{Cadets-E3}} & Real & \cellcolor{basegray}1.72 & \cellcolor{basegray}0.75 & \cellcolor{basegray}19.83 & \cellcolor{basegray}9.74 & & \cellcolor{basegray}0.00 & \cellcolor{basegray}0.50 & \cellcolor{basegray}0.00 & \cellcolor{basegray}0 \\
 & + ProvSyn & \cellcolor{downblue}1.94 & \cellcolor{downblue}0.84 & \cellcolor{downblue}25.41 & \cellcolor{downblue}35.24 & & \cellcolor{downblue}4.85 & \cellcolor{downblue}0.59 & \cellcolor{downblue}20.47 & \cellcolor{downblue}12.55 \\
\cmidrule(lr){1-11}
\multirow{2}{*}{\textbf{Theia-E3}} & Real & \cellcolor{basegray}0.11 & \cellcolor{basegray}0.47 & \cellcolor{basegray}-2.55 & \cellcolor{basegray}0.31 & & \cellcolor{basegray}0.22 & \cellcolor{basegray}0.44 & \cellcolor{basegray}-3.24 & \cellcolor{basegray}0.56 \\
 & + ProvSyn & \cellcolor{downblue}0.60 & \cellcolor{downblue}0.53 & \cellcolor{downblue}2.71 & \cellcolor{downblue}1.89 & & \cellcolor{downblue}0.44 & \cellcolor{downblue}0.53 & \cellcolor{downblue}1.99 & \cellcolor{downblue}1.41 \\
\cmidrule(lr){1-11}
\multirow{2}{*}{\textbf{ClearScope-E5}} & Real & \cellcolor{basegray}0.29 & \cellcolor{basegray}0.47 & \cellcolor{basegray}-3.27 & \cellcolor{basegray}0.41 & & \cellcolor{basegray}0.58 & \cellcolor{basegray}0.48 & \cellcolor{basegray}-1.44 & \cellcolor{basegray}0.82 \\
 & + ProvSyn & \cellcolor{downblue}0.97 & \cellcolor{downblue}0.61 & \cellcolor{downblue}7.93 & \cellcolor{downblue}2.50 & & \cellcolor{downblue}0.83 & \cellcolor{downblue}0.57 & \cellcolor{downblue}5.53 & \cellcolor{downblue}2.00 \\
\cmidrule(lr){1-11}
\multirow{2}{*}{\textbf{Theia-E5}} & Real & \cellcolor{basegray}1.16 & \cellcolor{basegray}0.73 & \cellcolor{basegray}15.56 & \cellcolor{basegray}12.51 & & \cellcolor{basegray}0.13 & \cellcolor{basegray}0.35 & \cellcolor{basegray}-14.90 & \cellcolor{basegray}0.08 \\
 & + ProvSyn & \cellcolor{upgreen}1.03 & \cellcolor{downblue}0.80 & \cellcolor{downblue}17.20 & \cellcolor{downblue}18.63 & & \cellcolor{downblue}0.30 & \cellcolor{downblue}0.60 & \cellcolor{downblue}4.08 & \cellcolor{downblue}2.29 \\
\cmidrule(lr){1-11}
\multirow{2}{*}{\textbf{OpTC-H201}} & Real & \cellcolor{basegray}32.24 & \cellcolor{basegray}0.72 & \cellcolor{basegray}671.63 & \cellcolor{basegray}15.61 & & \cellcolor{basegray}0.11 & \cellcolor{basegray}0.17 & \cellcolor{basegray}N/A & \cellcolor{basegray}0.01 \\
 & + ProvSyn & \cellcolor{downblue}32.61 & \cellcolor{basegray}0.72 & \cellcolor{downblue}1446.15 & \cellcolor{downblue}15.81 & & \cellcolor{downblue}5.12 & \cellcolor{downblue}0.55 & \cellcolor{downblue}293.88 & \cellcolor{downblue}1.53 \\
\cmidrule(lr){1-11}
\multirow{2}{*}{\textbf{OpTC-H501}} & Real & \cellcolor{basegray}6.10 & \cellcolor{basegray}0.70 & \cellcolor{basegray}N/A & \cellcolor{basegray}10.73 & & \cellcolor{basegray}0.15 & \cellcolor{basegray}0.26 & \cellcolor{basegray}-335.60 & \cellcolor{basegray}0.07 \\
 & + ProvSyn & \cellcolor{downblue}6.82 & \cellcolor{downblue}0.72 & \cellcolor{downblue}225.50 & \cellcolor{downblue}12.89 & & \cellcolor{downblue}0.86 & \cellcolor{downblue}0.46 & \cellcolor{downblue}-35.13 & \cellcolor{downblue}0.73 \\
\midrule
\midrule
 & & \multicolumn{4}{c}{\textbf{Magic}} & & \multicolumn{4}{c}{\textbf{Velox}} \\
\cmidrule(lr){3-6} \cmidrule(lr){8-11}
\textbf{Dataset} & \textbf{Source} & \textbf{F-Score} & \textbf{B.Acc.} & \textbf{MCC} & \textbf{DOR} & & \textbf{F-Score} & \textbf{B.Acc.} & \textbf{MCC} & \textbf{DOR} \\
\midrule
\multirow{2}{*}{\textbf{Cadets-E3}} & Real & \cellcolor{basegray}0.99 & \cellcolor{basegray}0.73 & \cellcolor{basegray}14.61 & \cellcolor{basegray}12.92 & & \cellcolor{basegray}0.00 & \cellcolor{basegray}0.50 & \cellcolor{basegray}-0.04 & \cellcolor{basegray}0 \\
 & + ProvSyn & \cellcolor{downblue}1.54 & \cellcolor{downblue}0.80 & \cellcolor{downblue}20.76 & \cellcolor{downblue}17.49 & & \cellcolor{downblue}3.48 & \cellcolor{downblue}0.51 & \cellcolor{downblue}6.43 & \cellcolor{downblue}7.88 \\
\cmidrule(lr){1-11}
\multirow{2}{*}{\textbf{Theia-E3}} & Real & \cellcolor{basegray}0.42 & \cellcolor{basegray}0.57 & \cellcolor{basegray}N/A & \cellcolor{basegray}2.00 & & \cellcolor{basegray}54.61 & \cellcolor{basegray}0.57 & \cellcolor{basegray}67.78 & \cellcolor{basegray}243.04 \\
 & + ProvSyn & \cellcolor{downblue}0.63 & \cellcolor{downblue}0.72 & \cellcolor{basegray}N/A & \cellcolor{downblue}12.46 & & \cellcolor{downblue}140.35 & \cellcolor{upgreen}0.55 & \cellcolor{downblue}151.65 & \cellcolor{downblue}1934.86 \\
\cmidrule(lr){1-11}
\multirow{2}{*}{\textbf{ClearScope-E5}} & Real & \cellcolor{basegray}0.63 & \cellcolor{basegray}0.48 & \cellcolor{basegray}-1.20 & \cellcolor{basegray}0.88 & & \cellcolor{basegray}48.19 & \cellcolor{basegray}0.52 & \cellcolor{basegray}49.25 & \cellcolor{basegray}205.28 \\
 & + ProvSyn & \cellcolor{downblue}0.86 & \cellcolor{downblue}0.61 & \cellcolor{downblue}9.32 & \cellcolor{downblue}15.30 & & \cellcolor{downblue}174.76 & \cellcolor{downblue}0.59 & \cellcolor{downblue}174.48 & \cellcolor{downblue}751.84 \\
\cmidrule(lr){1-11}
\multirow{2}{*}{\textbf{Theia-E5}} & Real & \cellcolor{basegray}0.30 & \cellcolor{basegray}0.69 & \cellcolor{basegray}7.45 & \cellcolor{basegray}44.23 & & \cellcolor{basegray}54.79 & \cellcolor{basegray}0.51 & \cellcolor{basegray}120.37 & \cellcolor{basegray}11192.40 \\
 & + ProvSyn & \cellcolor{downblue}0.39 & \cellcolor{downblue}0.76 & \cellcolor{downblue}10.07 & \cellcolor{downblue}79.14 & & \cellcolor{downblue}56.34 & \cellcolor{basegray}0.51 & \cellcolor{downblue}170.24 & \cellcolor{downblue}$\infty$ \\
\cmidrule(lr){1-11}
\multirow{2}{*}{\textbf{OpTC-H201}} & Real & \cellcolor{basegray}1.67 & \cellcolor{basegray}0.27 & \cellcolor{basegray}-1882.50 & \cellcolor{basegray}0.13 & & \cellcolor{basegray}0.68 & \cellcolor{basegray}0.50 & \cellcolor{basegray}3.61 & \cellcolor{basegray}21.48 \\
 & + ProvSyn & \cellcolor{basegray}1.67 & \cellcolor{downblue}0.28 & \cellcolor{downblue}-915.75 & \cellcolor{downblue}0.15 & & \cellcolor{downblue}0.69 & \cellcolor{basegray}0.50 & \cellcolor{downblue}4.82 & \cellcolor{downblue}38.01 \\
\cmidrule(lr){1-11}
\multirow{2}{*}{\textbf{OpTC-H501}} & Real & \cellcolor{basegray}0.42 & \cellcolor{basegray}0.26 & \cellcolor{basegray}N/A & \cellcolor{basegray}0.11 & & \cellcolor{basegray}2.65 & \cellcolor{basegray}0.50 & \cellcolor{basegray}13.77 & \cellcolor{basegray}331.65 \\
 & + ProvSyn & \cellcolor{downblue}2.04 & \cellcolor{downblue}0.63 & \cellcolor{basegray}N/A & \cellcolor{downblue}3.04 & & \cellcolor{downblue}2.66 & \cellcolor{basegray}0.50 & \cellcolor{downblue}21.07 & \cellcolor{downblue}994.95 \\
\bottomrule
\end{tabular}
}
\end{table*}

To demonstrate the practical utility of \Sys, we focus on APT attack detection using four representative systems: NodLink~\cite{li2023nodlink}, Flash~\cite{rehman2024flash}, Magic~\cite{jia2024magic}, and Velox~\cite{bilot2025sometimes}. All systems are reproduced using the open-source PIDSMaker~\cite{bilot2026pidsmaker,bilot2025sometimes} framework. Our evaluation focuses on whether augmenting the training set with synthesized provenance graphs enhances detection performance. To this end, we incorporate synthetic graphs, as graph communities, into the training data.
The detection performance is presented in Table~\ref{tab:ids}. Based on the results, we have the following four key findings.

\PP{Detection Performance Enhancement.}
The integration of \Sys data consistently strengthens detection metrics across various baselines, demonstrating its efficacy in both high-performing and low-performing scenarios. For instance, when a model already exhibits reasonable performance, such as NodLink on the Cadets-E3 dataset, the B.Acc. increases from $0.75$ to $0.84$ and the F-Score improves from $1.72 \times 10^{-3}$ to $1.94 \times 10^{-3}$. More importantly, the augmentation provides a critical correction in cases where the initial performance is near or below random chance. A prominent example is observed with the Flash algorithm on the OpTC-H201 dataset, where the B.Acc. rises significantly from $0.17$ to $0.55$. Similar improvements are seen in the OpTC-H501 dataset for Flash, where the B.Acc. progresses from $0.26$ to $0.46$, effectively rescuing the model from a state of total detection failure.

\PP{Discriminative Reliability and Diagnostic Utility.}
The impact of \Sys is further evidenced by the improvement in MCC and DOR, which reflect the model's correlation with the ground truth and its diagnostic strength. The MCC values are consistently pulled toward positive territory, particularly in instances where the baseline models initially yield negative results, such as NodLink on Theia-E3 moving from $-2.55$ to $2.71$ and Flash on ClearScope-E5 shifting from $-1.44$ to $5.53$. This indicates a fundamental shift from biased or inverse classification to a correct positive correlation. Simultaneously, the DOR increases across all experiments, often by several orders of magnitude. For example, the DOR for Velox on Theia-E3 jumps from $243.04$ to $1934.86$, and on Theia-E5, it reaches $\infty$. These results suggest that the diagnostic efficiency for identifying attack samples is substantially sharpened through the augmented training set.

\PP{Enhancement Mechanism Analysis.} The overall performance uplift across different dataset--IDS combinations can be attributed to the mitigation of data imbalance and the diversity introduced by \Sys synthesized samples. By incorporating a wider variety of benign provenance graphs into the training set, the models can cover previously underrepresented system behaviors and edge-case scenarios. This expanded representation of the benign space helps the detection algorithms to more accurately define the boundaries of normal activity. Consequently, the models become less prone to overfitting on limited real-world benign data, enabling them to better distinguish subtle malicious deviations from legitimate but rare system actions, thereby improving the robustness of APT detection.

\subsection{Overhead Analysis}

We analyze the time and memory requirements of \Sys. The most time- and resource-intensive components of the \Sys framework are the heterogeneous graph synthesis module and the text attribute synthesis module.

For the graph structure synthesis module, we use 16,000 samples for training, with a batch size of 32 and 3,000 training epochs. The average training time is 18 hours, with a GPU memory usage of 0.84 GB. Inference for generating 1,000 graphs takes 28.09 seconds and consumes 0.38 GB of memory.
For the text attribute synthesis module, we adopt Llama3.2-3B-Instruct as the base model, applying the LoRA algorithm with 4-bit quantization. The FastLanguageModel library is used to accelerate training. The training consists of 60 steps, takes an average of 109 seconds, and requires a peak memory of 3.44 GB. During inference, generating text attributes for a single graph takes an average of 80 seconds. In our experiments, we deploy 8 fine-tuned LLMs in parallel to synthesize 1,000 graphs, completing the process in 2.78 hours. The total GPU memory required for the 8 models is 28 GB. This analysis demonstrates that \Sys is a time- and memory-efficient framework for provenance graph synthesis.

\section{Related Work}
\PP{Provenance-based Intrusion Detection System (\ac{pids}).} Provenance-based detection methods can be divided into three main categories~\cite{zengy2022shadewatcher}: rule-based, statistics-based, and learning-based techniques. Rule-based methods~\cite{hassan2020tactical,milajerdi2019holmes,hossain2020combating} rely on predefined heuristics derived from known attack patterns to identify malicious activities. Statistics-based~\cite{hassan2019nodoze,wang2020you,liu2018towards} approaches analyze deviations in provenance graph elements by measuring their statistical anomalies. Learning-based techniques employ deep learning models to capture either normal system behavior~\cite{han2020unicorn,wang2022threatrace} or malicious patterns~\cite{liu2019log2vec,zengy2022shadewatcher,li2021hierarchical}, framing APT detection as either a classification~\cite{li2021hierarchical} or anomaly detection task~\cite{zengy2022shadewatcher,han2021sigl}. These learning-based methods further branch into sequence-based approaches~\cite{liu2019log2vec}, which focus on execution workflows, and graph-based~\cite{han2020unicorn,wang2022threatrace,zengy2022shadewatcher,han2021sigl,li2021hierarchical,wang2025rr} methods, which use graph neural networks to model entity relationships and detect anomalies.

\PP{LLM-driven Synthetic Data Generation.}
LLM-based synthetic data generation methods can be categorized into three main types: prompt engineering-based approaches~\cite{gilardi2023chatgpt,gunasekar2023textbooks,sudalairaj2024lab,liu2024best,yoo2021gpt3mix,li2023textbooks,ye2022progen}, which utilize task descriptions, condition-value prompts, and in-context examples to steer generation; multi-step generation-based approaches~\cite{wei2022chain,gao2023retrieval,ding2023enhancing,wang2022self,honovich2022unnatural}, which break down complex tasks either at the sample or dataset level; and knowledge-enhanced approaches~\cite{pan2024unifying,chen2024mindsearch,ji2021survey}, which incorporate external knowledge such as knowledge graphs or web sources to improve the factual quality of outputs. The evaluation of synthetic data includes both direct metrics, which measure faithfulness~\cite{lee2022factuality} and diversity~\cite{yu2023large}, and indirect metrics, which assess performance on downstream tasks through benchmark testing~\cite{sun2023principle} and open-ended evaluation using human~\cite{he2023annollm} or model-based~\cite{xu2023wizardlm} methods.

%% file: discussion.tex
\section{Discussion and Limitations}
\PP{Temporal Modeling.} A provenance graph is not only a directed heterogeneous graph with textual attributes, but also a temporal graph in which each event is associated with a timestamp indicating its occurrence time in the system. This temporal information has recently been leveraged by downstream intrusion detection systems~\cite{rehman2024flash} to better capture the patterns of benign and anomalous behaviors. However, in \Sys, we do not synthesize explicit timestamps for events. Instead, we approximate the temporal order by relying on the sequence of directed edges. Future work may explore modeling temporal dynamics using temporal graph generation networks~\cite{zhang2020tg}, or designing strategies that leverage LLMs to generate realistic timestamps, thereby enabling more comprehensive and temporally coherent provenance graph synthesis.

\PP{Graph Scale.} In \Sys, the generated provenance graphs are smaller in scale compared to real-world provenance graphs. To enable data augmentation, we embed multiple synthetic graphs as distinct communities within large real-world provenance graphs. The results show that this approach effectively mitigates the imbalance of data and improves the accuracy of the detection models trained on the augmented datasets. Future work can focus on scaling up synthetic provenance graphs by enhancing graph generation models for large-scale synthesis and exploring graph merging algorithms that combine smaller graphs into larger ones while preserving structural fidelity.

\PP{Application Scenarios.} To date, the graphs synthesized by \Sys have demonstrated utility within the domain of provenance graph datasets and for detection algorithms targeting APT attacks. However, \Sys is a general framework capable of synthesizing heterogeneous high-fidelity graphs with textual attributes. This gap between \Sys's inherent capabilities and current applications presents an opportunity for future work: expanding the exploration of \Sys's capabilities to encompass a wider range of application scenarios and downstream algorithms.

%% file: appendix.tex
\section{Entity and Event Distribution}
\label{sec:entity-event-distribution}

\begin{figure*}[htbp]
    \centering
    \includegraphics[width=\textwidth]{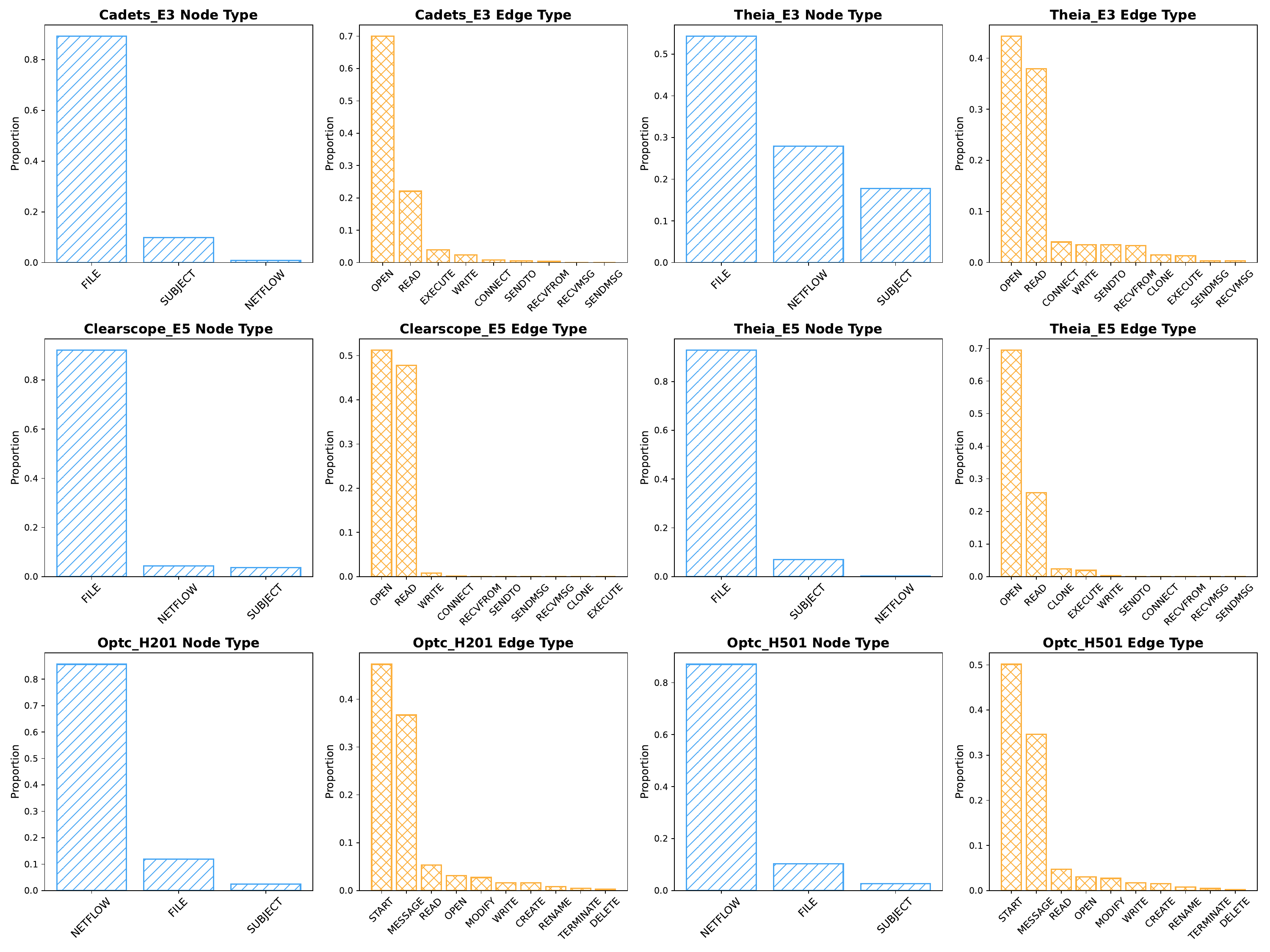}
    \caption{Entity Type and Event Type Distribution in Provenance Dataset.}
    \label{fig:label_stats}
\end{figure*}

We present the distributions of \textit{entity type} and \textit{event type} across six commonly-used public provenance datasets in Figure \ref{fig:label_stats}. The distributions exhibit a typical long-tail pattern and reveal significant data imbalance.

\section{Schema Rules for Decoding}
\label{sec:schema}
For each dataset, we define a decoding schema that constrains valid edge types between source and destination node pairs, as summarized in Table~\ref{tab:schema}. Derived from ground-truth provenance graphs, these schemas are used to mask invalid relations during decoding.

\begin{table*}[t]
\centering
\caption{Dataset-specific schemas}
\label{tab:schema}
\resizebox{\textwidth}{!}{%
\begin{tabular}{c c c}
\toprule
\textbf{Dataset} & \textbf{Node Pair (Src, Dst)} & \textbf{Permissible Event Types} \\
\midrule
\multirow{5}{*}{DARPA E3/E5} & (subject, subject) & read, write, open, connect, recvfrom, sendto, clone, sendmsg, recvmsg \\
                             & (subject, file)    & write, connect, sendmsg, sendto, clone \\
                             & (subject, netflow) & write, sendto, connect, sendmsg \\
                             & (file, subject)    & read, open, recvfrom, execute, recvmsg \\
                             & (netflow, subject) & open, read, recvfrom, recvmsg \\
\midrule
\multirow{5}{*}{OpTC}        & (subject, subject) & create, open, terminate \\
                             & (subject, file)    & create, delete, modify, rename, write \\
                             & (file, subject)    & read \\
                             & (subject, netflow) & message, start \\
                             & (netflow, subject) & message, open, start \\
\bottomrule
\end{tabular}%
}
\end{table*}

\section{Baseline LLM Prompt}
\label{sec:prompt}
We present in Figure~\ref{fig:prompt} the prompt used to instruct LLMs in generating provenance logs on Cadets-E3 dataset. In our experiments, the Examples section contains 20 examples. For presentation brevity, only 5 examples are shown.

\section{Hyperparameters Setting}
\label{sec:hyper}

\begin{figure*}[htbp] 
    \centering
    \includegraphics[width=\textwidth]{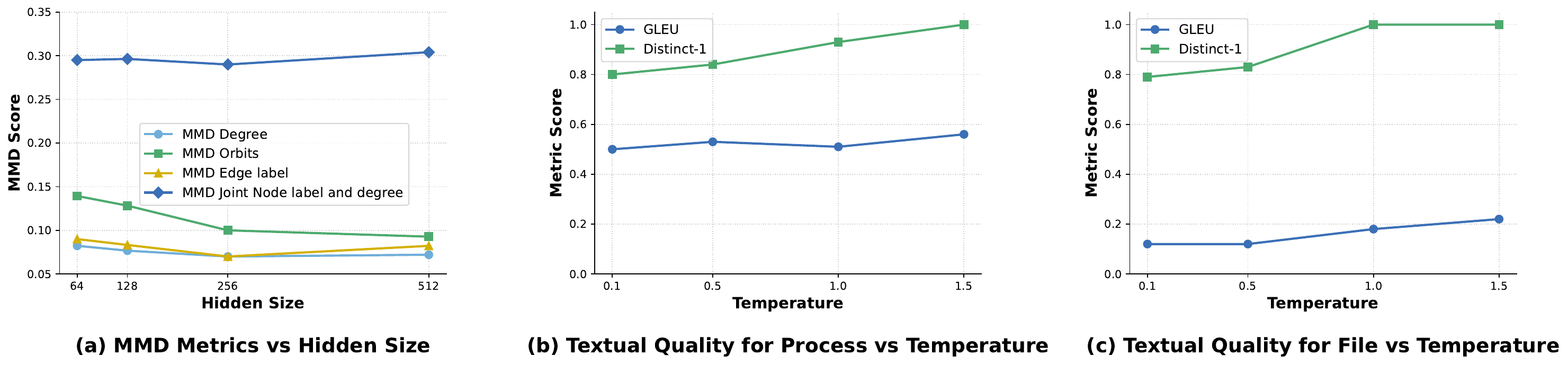}
    \caption{Hidden size setting and inference temperature setting.} 
    \label{fig:hyperparameter} 
\end{figure*}

\PP{Hidden size} is a key parameter of LSTM for graph structure synthesis, as it determines the capacity of the model to capture structural dependencies in sequential node and edge generation. We conducted experiments on the Theia-E3 dataset and evaluated structural fidelity using five MMD-based metrics in the validation set. The results in Figure \ref{fig:hyperparameter} (a) show that a hidden size of 64 yields significantly lower fidelity compared to the other settings. For hidden sizes of 128, 256, and 512, the MMD values remain at a comparable level, with 256 performing slightly better than 128 and 512. Considering both performance and computational cost, we select 256 as the hidden size for \Sys.

\PP{Inference temperature} is a key parameter in LLM-based inference, controlling the randomness of generated text. Higher temperatures yield more diverse outputs, while lower temperatures produce more deterministic results. We conducted experiments on the Cadets-E3 dataset using temperature values of 0.1, 0.5, 1.0, and 1.5, with all other inference parameters held constant. Evaluation focused on the quality of text generated for process and file node attributes. As shown in Figure \ref{fig:hyperparameter} (b) and (c), the best performance is achieved when the temperature is set to 1.5. Based on these findings, we selected 1.5 as the final inference temperature for \Sys.

\section{Ablation Study}
\label{sec:ablation}
\begin{figure*}[htbp]
    \centering
    \includegraphics[width=\textwidth]{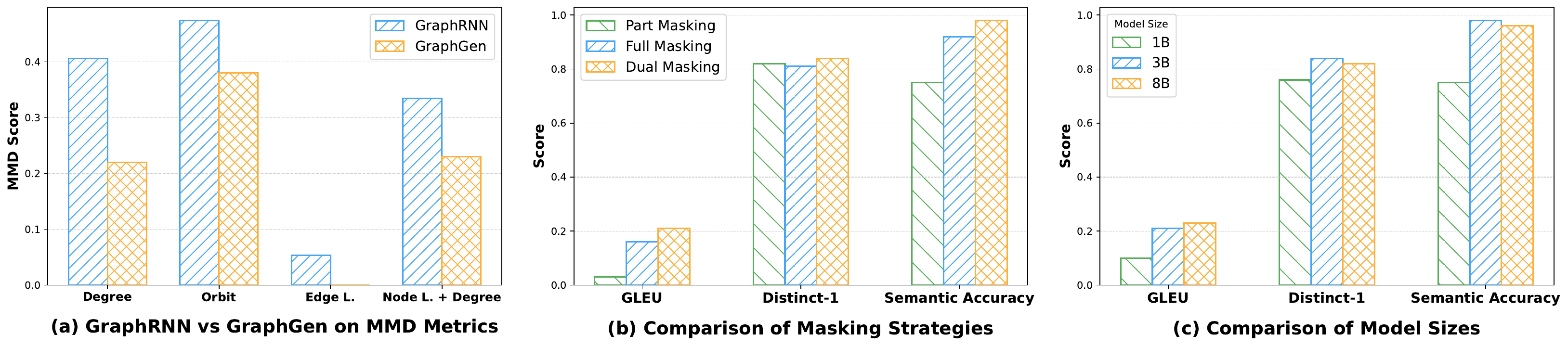}
    \caption{Ablation Study.} 
    \label{fig:ablation} 
\end{figure*}

\PP{Alternative models to GraphGen.} For heterogeneous graph structure generation, we adopt the design of GraphGen, which encodes graphs as DFS code sequences and employs an LSTM for training and inference. To validate the effectiveness of this design, we compare it with alternative models, including the widely used GraphRNN~\cite{you2018graphrnn}. In our experiments, we evaluate the graphs generated by GraphRNN and GraphGen using four MMD metrics. As shown in Figure \ref{fig:ablation}(a), the graphs generated by GraphGen consistently achieve lower MMD scores across all MMD metrics, indicating better fidelity compared to those generated by GraphRNN. These results demonstrate the effectiveness of using GraphGen for heterogeneous graph generation in \Sys.

\PP{Masking strategies.} For text attribute generation, we employ two masking strategies to construct the training set, corresponding to two different input sequences during inference. To evaluate the effectiveness of the masking design, we perform ablation experiments by disabling one of the strategies and using only the other. Model performance is then evaluated using textual and semantic metrics. The experiments are conducted on the ClearScope-E5 dataset, with the number of training samples fixed at 20,000 for both the single-strategy and dual-strategy settings.
In the results shown in Figure \ref{fig:ablation}(b), we observe that using only the Full Masking strategy yields better performance than using only the Part Masking strategy. This may be because the Part Masking strategy provides partial node names within the sequence, offering more cues for the LLM to leverage. As a result, a model trained solely with Part Masking performs poorly when tested on Full Masking sequences, which contain fewer clues. In contrast, using both strategies together achieves better performance than either alone, as it exposes the model to a wider range of sequence types that may appear during inference. These results demonstrate the effectiveness of our proposed approach that combines both masking strategies.

\PP{LLM parameter size.} In our experiments on text attribute generation, we employ the Llama-3.2-3B-Instruct model as the base model for training. We also compare its performance against models of different sizes, including Llama-3.2-1B-Instruct and Llama-3.1-8B-Instruct. All models are trained on the same training datasets and evaluated using textual and semantic metrics. The results, presented in Figure \ref{fig:ablation}(c), demonstrate that the 3B model achieves the highest scores on the Distinct-1 and semantic accuracy. In contrast, the 8B model shows marginally better performance on the GLEU. Given that the advantage of the 8B model is not substantial and considering computational resource constraints, we select the 3B model as the base for synthesizing text attributes. In practice, \Sys introduces a general training strategy to leverage LLMs for synthesizing text attributes, where the choice of the base model involves balancing the complexity of the dataset, the application requirements and the limitations of computational resources.

\begin{figure*}[t]
\centering
\begin{lstlisting}
[Task Description] You are an audit log generator. Carefully analyze the audit log records provided in [Examples] to understand their structure, semantics, and temporal patterns. Based on this understanding, generate 50 unique audit log records that are logically consistent, structurally valid, and realistic. The generated logs should reflect plausible system activities, preserve coherent relationships between entities and events, and follow realistic ordering observed in the examples.

[Examples]

[
 {
    "event_type": "EVENT_OPEN",
    "source_node_type": "file",
    "source_node_name": "kernel_mem.txt",
    "target_node_type": "subject",
    "target_node_name": "vmstat"
  },
  {
    "event_type": "EVENT_READ",
    "source_node_type": "file",
    "source_node_name": "None",
    "target_node_type": "subject",
    "target_node_name": "pickup"
  },
  {
    "event_type": "EVENT_WRITE",
    "source_node_type": "subject",
    "source_node_name": "head",
    "target_node_type": "file",
    "target_node_name": "top_procs.txt"
  },
  {
    "event_type": "EVENT_CONNECT",
    "source_node_type": "subject",
    "source_node_name": "master",
    "target_node_type": "file",
    "target_node_name": "None"
  },
  {
    "event_type": "EVENT_EXECUTE",
    "source_node_type": "file",
    "source_node_name": "/usr/bin/top",
    "target_node_type": "subject",
    "target_node_name": "top"
  },
  ......
]

[Output] Return the logs in JSON format. Only output the JSON.
\end{lstlisting}
\caption{Prompt for LLM Provenance Data Generation on Cadets-E3}
\label{fig:prompt}
\end{figure*}